# Large-Area Spatially Ordered Mesa Top Single Quantum Dots: Suitable Single Photon Emitters for On-Chip Integrated Quantum Information Processing Platforms

Qi Huang, Lucas Jordao, Siyuan Lu[*], Swarnabha Chattaraj[†], Jiefei Zhang[†], & Anupam Madhukar[‡]

*Nanostructure Materials and Devices Laboratory, University of Southern California, Los Angeles, CA 90089-0241*

**Date: Dec. 21, 2023**

## Abstract

Realization of the long sought on-chip scalable photonic quantum information processing networks has been thwarted by the absence of *spatially-ordered and scalable on-demand single photon emitters* with emission figures-of-merit exceeding the required thresholds across large numbers. The positioning must meet the required degree of accuracy that enables fabricating their interconnection to create the desired functional network. Here we report on the realization of *large-area* spatially-ordered arrays of mesa-top single quantum dots (MTSQDs) that are demonstrated [1] to be on-demand single photon emitters with characteristics that meet the requirements for implementing quantum photonic circuits/platforms aimed at quantum key distribution, linear optical quantum computing, simulations of quantum many-body problems, and metrology/sensing. The reported GaAs/InGaAs/GaAs MTSQD arrays, grown via SESRE (substrate-encoded size-reducing epitaxy) are in multiple arrays of up to 100×100 with 5μm pitch, across a centimeter radius area. We show illustrative large-area images of the emission intensity (brightness) and color-coded wavelength distribution exhibiting ~3.35nm standard deviation. Scanning transmission electron microscopy shows a remarkable control on the QD location to within ~3nm accuracy laterally and ~1nm vertically. The primary remaining challenge is the control on the uniformity of the currently wet-chemically etched *as-patterned* nanomesa lateral size across the substrate, a surmountable technical issue. Thus, SESRE offers the most promising approach to realizing on-chip scalable *spatially-ordered* arrays of on-demand bright single quantum emitters meeting the figures-of-merit required for on-chip fully integrated quantum photonic circuit platforms-monolithic (such as based upon AlGaAs on insulator) or hybrid that leverage the silicon-on-insulator (SOI) photonic integrated circuit (PIC).

---

[*] Currently at Cruise LLC. San Francisco. California. 94107, United States.
[†] Currently at Argonne National Laboratory. Illinois. 60439, United States.
[‡] Author to whom correspondence should be addressed. Electronic mail: madhukar@usc.edu.

**I. Introduction**

Three basic requirements for realizing on-chip integrated quantum photonic networks / circuits for quantum information processing (QIP) relate to the photon emitters: these are (1) *spatial locations known with adequate accuracy; (2) on-demand emission; (3) sufficiently uniform spectral figures-of-merit exceeding established thresholds* across the large arrays of bright on-demand quantum emitters. The lack of such quantum emitter arrays has hitherto only allowed proof-of-principle demonstrations of minimal on-chip function from a few (typically <10)) interconnected quantum emitters [2]. Here we report the scaling up of our recently demonstrated [1] unique class of epitaxial semiconductor quantum dots, dubbed mesa top single quantum dots (MTSQDs), that meet all the above noted requirements, thus opening the path to exploration of scalable on-chip quantum information systems.

The distribution of the degree of overall complexity embodied in a network (such as for computation, secure communication, metrology) across identified simpler functional building blocks is the essential architecture of all information processing networks ranging from the evolution-driven "brains" of increasingly complex organisms to man-made architectures such as the sophisticated very large-scale integration (VLSI) based digital chips. Analogous to electronic VLSI technology is the currently conceived photonic integrated circuits *that include, on-chip,* the requisite on-demand quantum emitters, along with all the needed emitted-photon manipulation circuitry (filtering, frequency conversion, memory, beam-couplers, etc., as-needed) as well as efficient photon number counting detectors. Silicon-on-insulator (SOI) based platforms provide much of the desired photon manipulation circuitry-- particularly in the communication C-band (~1550nm) wavelength leveraged by the existing global infrastructure of low photon loss optical fiber network for classical communication—and are becoming increasingly sophisticated in handling single and few photon signals [3] but lack (i) on-chip deterministic and efficient / bright quantum emitters and (ii) adequately low-noise single photon number counting detectors. For quantum information processing, invariably entangled matter and light qubits with controlled / tailorable interaction between them (light-matter interface) are necessary. Faithful conversion, transport, manipulation, and measurement of information cycling between entangled matter and photon qubits within the coherence time of the network is thus the task of a QIP platform. Photons, by virtue of their weak interaction with environment and each other, are the ideal choice for carrying, faithfully, quantum information between matter qubits over distances ranging from on-chip (microns) to connecting, through free space, any two points on earth thousands of kilometers apart. Dubbed flying qubits, single and controlled number of entangled photon states generated, on-demand, are thus an essential quantum resource.

Semiconductor epitaxial quantum dots, through the work of many researchers over nearly two decades [see the comprehensive compilations in [4, 5]], have been demonstrated to be the only on-demand solid-state quantum emitters that are also bright and have nearly unity internal quantum efficiency—characteristics essential for the best quantum information processing platforms. As a guide, Figure. 1(a) and (b) capture, respectively, the acknowledged relationships (broken lines) between single photon purity and generation probability [6] and two-photon HOM (Hong-Ou-Mandel) visibility and photon detection probability [7] that delineates the required regime (the right side) of characteristics suitable for QIP. The red star denotes epitaxial quantum dots and captures the fact that these are the only demonstrated single photon sources with values of these figures-of-merit above the broken black lines. Indeed, these values reach near unity, i.e. in the required region of the parameter space for efficient QIP. What has been lacking until now is realizing epitaxial quantum dots that are in scalable spatially ordered locations and simultaneously satisfy the above noted spectral requirements across the array. In the remainder of this paper we provide, most briefly, the encouraging findings of our continued efforts to establish the limits of our SESRE approach to realizing spatially ordered epitaxial SQDs in scalable large arrays.

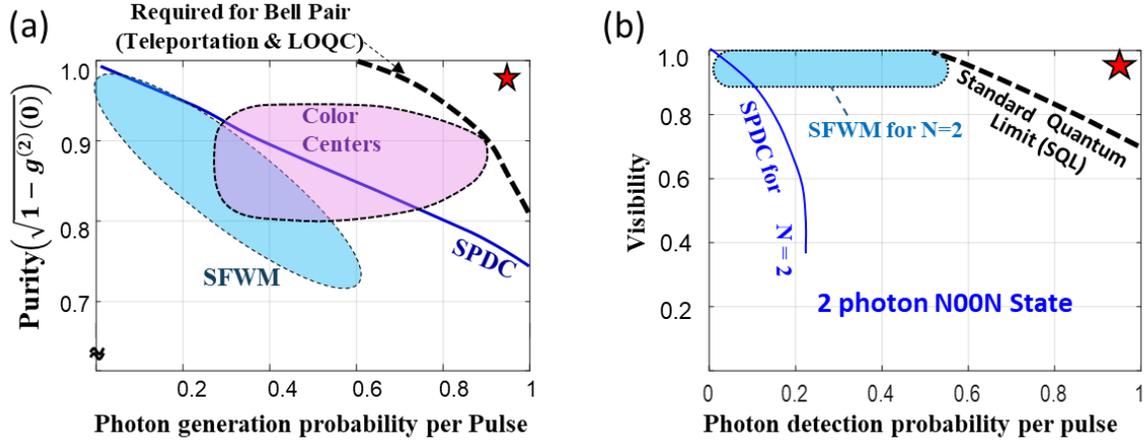

**Figure 1.** Shows the required performance regime (above the broken lines) for a single photon source to serve quantum information processing platforms [Adapted from ref. 1]: (a) purity versus photon generation probability per pulse for linear optical quantum computing (broken line from ref. 6); (b) visibility versus photon detection probability per pulse for 2-photon NOON state generation for quantum sensing at the standard quantum limit (broken line from ref. 7). Semiconductor epitaxial quantum dots (red star) are the only solid-state quantum emitters that *individually* have been demonstrated to meet these criteria [4,5].

## II. Suitable Quantum Emitters in Large Arrays

The commonly used epitaxial quantum dots—the lattice-mismatch strain-driven and spontaneously formed self-assembled quantum dots (SAQDs)-- have been demonstrated to satisfy the criteria depicted in Figure 1. However, their random spatial location and large (~40nm) emission wavelength nonuniformity owing to the fluctuations in size, shape, and composition have prevented them from being suitable for creating networks of coupled quantum emitters. The class of quantum dots dubbed mesa-top single quantum dots (MTSQDs) have, however, been demonstrated [1] to satisfy the photon characteristics as required in Figure 1 and are located in designed spatially ordered arrays. We have scaled these reported 5×8 arrays [1] to 100×100 with the 10,000 MTSQDs distributed over 0.5mm × 0.5mm area and such arrays spread over a growth area of ~2cm diameter, the area over which our molecular beam epitaxy growth system has 1% uniformity of flux distribution during growth.

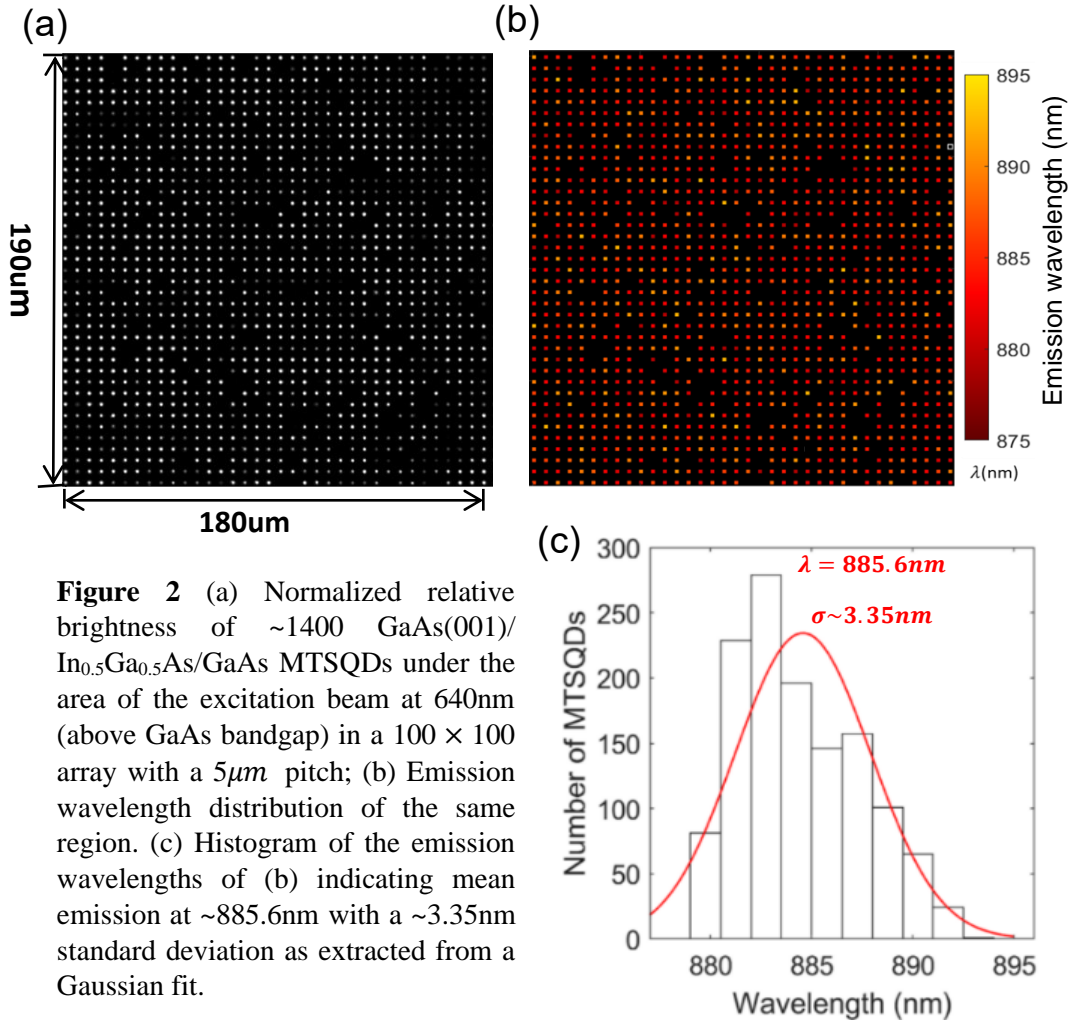

**Figure 2** (a) Normalized relative brightness of ~1400 GaAs(001)/In$_{0.5}$Ga$_{0.5}$As/GaAs MTSQDs under the area of the excitation beam at 640nm (above GaAs bandgap) in a $100 \times 100$ array with a $5\mu m$ pitch; (b) Emission wavelength distribution of the same region. (c) Histogram of the emission wavelengths of (b) indicating mean emission at ~885.6nm with a ~3.35nm standard deviation as extracted from a Gaussian fit.

Figure 2 shows emission from ~1400 MTSQDs taken from an *as-grown* $100 \times 100$ array with a $5\mu m$ pitch. To enhance the collection efficiency for measurement purposes, the MTSQDs are grown on substrates with appropriately designed buried distributed Bragg mirror (DBR). The shown area is limited by the width of the excitation beam derived from a broad-spectrum continuous lamp source. The excitation wavelength is 640nm (above GaAs bandgap) and the power is 0.2W/cm$^2$, safely in the single photon emission regime. The emission intensities, shown in Figure 2(a), are normalized to the same incident photon flux and thus represent relative brightness. The emission wavelengths of these same MTSQD region of the array are mapped employing a tunable filtering system allowing a spectral resolution of ~1.5nm for parallel measurements over large areas [9]. A color-coded emission wavelength image obtained at 1.6nm spectral resolution is shown in Figure 2(b). The histogram of the spectral distribution over these ~1400 MTSQDs covering an ~35000$\mu m^2$ area is shown in Figure 2(c). A Gaussian fit gives a ~3.35nm standard deviation for the emission wavelength across such an area. This remarkably narrow spectral emission distribution from SQDs without any concerted attempt to control the fluctuations in the *as-patterned* starting nanomesa lateral size distribution (wet chemical etching is employed), is highly encouraging. The spectral nonuniformity can be significantly reduced by improved control on the starting nanomesa size fluctuation. Additionally, employing binary materials (InAs/GaAs and GaAs/AlAs) for the QD/barrier regions eliminates alloy disorder scattering and significantly lowers the MTSQD emission nonuniformity—down to ~1.8nm for GaAs/InAs/GaAs-- as we have demonstrated earlier [10]. The overwhelming majority of the MTSQDs in the array emit within the range of well-established technologies for local tuning such as via the Stark shift to bring two or more MTSQDs into resonance as needed for creating interference and cluster states—a basic quantum resource.

The employed absolute incident photon density of 0.2W/cm² is in the linear response regime and corresponds to a very low ~4.3 % of the saturation intensity of the brightest SQDs. As we reported earlier, the internal quantum efficiency of our MTSQDs is nearly unity [1] and with improved material quality the lowest power density required to ensure on-demand single photon emission behavior will continually be reduced. Currently, the MBE growth chamber condition gives materials in which the spectral diffusion, combined with the alloy disorder scattering inherent to alloy InGaAs QDs, limit the typical PL linewidth to be ~30$\mu$eV [1; see appendix]. The typical neutral exciton decay lifetime ($T_1$) for the MTSQDs can range from ~1ns to much shorter ~0.3ns (without Purcell enhancement) as the SESRE approach allows control on the shape and size of the MTSQD, thus enabling realizing large-volume SQDs with corresponding large oscillator strengths of ~27 (compared to the typical island quantum dot value of ~9) [11]. For the highest quality materials obtained in MBE, reaching the intrinsic life-time limited linewidth is not an intrinsic limitation of the SESRE methodology but only the condition of the MBE growth system and chemical purity of the sources.

Besides the large (100 × 100) MTSQD arrays with a 5 $\mu m$ pitch, these grown samples also contain regions of "sparse" arrays with the pitch varying between 20$\mu$m and 200$\mu$m in one direction and between 10$\mu$m and 20$\mu$m in the orthogonal direction. Figure 3 (uppermost grey region) shows an illustrative measured intensity image of the MTSQDs in this "sparse" region of the chip. The schematic of the whole figure illustrates the built-in potential for lift-off and transfer of such arrays to a co-designed silicon photonics platform to explore hybrid integrated on-chip QIP systems.

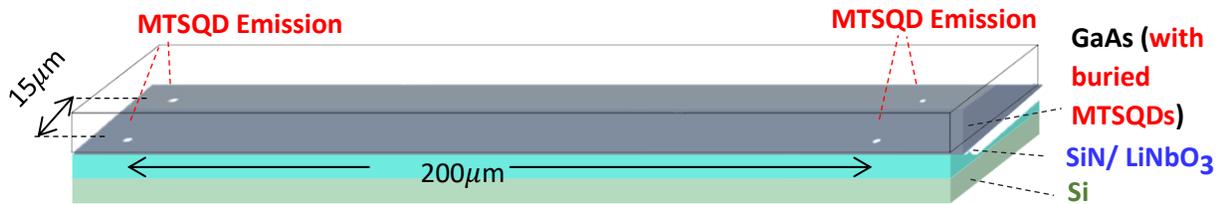

**Figure 3**. Uppermost grey region shows the measured intensity from four MTSQDs in a 5×20 array with a pitch of 200$\mu$m and 15$\mu$m, respectively. These sparse MTSQD regions are part of the same growth as the 100×100 MTSQD arrays of Figure 2. The schematically indicated underlying regions are to indicate potential hybrid integration of the MTSQDs with co-designed silicon photonics platforms.

We note that the large regions (containing ~10,000 to 1,000,000) of MTSQDs statistically contain subsets involving hundreds to thousands of MTSQDs emitting within ~0.2nm and thus suitable for fabricating the simplest of building blocks (BBs) such as the Mach-Zehnder Interferometer (MZI), that, when appropriately interconnected on-chip, enable realization of functional quantum photonic circuits. Indeed, as introduced by Dowling et al.[12], suitably modified MZI-based BBs constitute the "Rosetta stone" that enables a conceptually unifying view for implementing photonic networks across the gamut of quantum computing, communication, and metrology. The implementation can be either monolithic as in Figure 4(a) on a platform such as the AlGaAs on SOI [13] or via transfer-printing based hybrid integration of the MTSQD based directed single or entangled pair photon source arrays with silicon photonics-based chips [14] that provide all the desired emitted photon manipulation functional components as illustrated in Figure 4(b). This, much awaited step towards converting quantum optics to quantum photonic technology, is now made possible by MTSQD arrays as briefly addressed below.

Specifically, Figure 4(a) indicates that the as-grown sparse regions with long pitch along one direction are well-suited for fabricating MTSQDs embedded in photonic 2D crystal cavity and waveguide connecting seamlessly to continuum ridge waveguides thereby enabling (1) on-demand *directed* single photon sources in designed spatial arrays and (2) fabrication of inter-connected MTSQD networks to create hierarchical multi-photon entangled states. As with the N×N large arrays, these "sparse" regions with "L-shaped" local configurations, reveal triplets of as-grown MTSQDs emitting within the large-area measurement resolution limit of 0.2nm (~300 $\mu$eV).

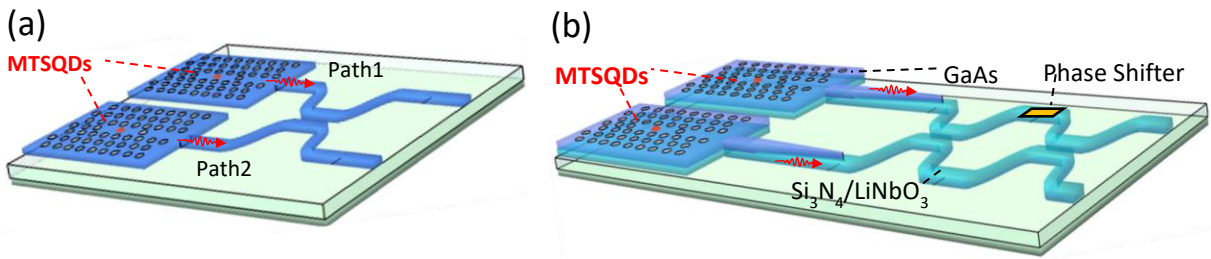

**Figure 4.** Panel (a) depicts schematically the coupled basic building blocks comprising the MTSQD embedded in a 2D photonic crystal cavity and WG that seamlessly dovetails into continuum ridge WG and on to a directional coupler (beam splitter). This provides the simplest configuration of on-chip directionally-emitting on-demand single photon sources coupled into a Mach-Zehnder interferometer—the "Rosetta stone" of QIP systems [12]. Implementing a network based upon such a building block with the MTSQDs residing on GaAs-on-insulator (GOI) substrate [13] will enable exploration of "monolithically" integrated on-chip platforms. Panel (b) depicts the same directional single photon emitting MTSQD building block and MZI interferometer-based network "hybrid" integrated via transfer to silicon photonics based emitted photon manipulation network [14].

## III. Incorporating MTSQD Arrays into Silicon Photonics Platform:

Silicon photonics is a mature technology for manipulation of light in the optical regime even though it is not efficient for generation of light. Nevertheless, exploiting the third order nonlinearity, spontaneous four wave mixing (SFWM) is being used to generate, probabilistically, pairs of photons that are subsequently processed into signal and heralded single photons, the latter for use as single photon sources needed for linear optical quantum computing. While the use of SFWM in silicon photonics based LOQC (linear optical quantum computing) is enabling unprecedented progress in creating large-scale photonic circuits [15], overcoming the non-deterministic and inefficient nature of SFWM impose the undesirable burden of incorporating extensive compensative on-chip circuitry [16]. This price would not need to be paid if deterministic, bright, and scalable SPS sources were available to hybrid integrate with the ensuing, much simplified, silicon circuitry as symbolically indicated in Figure 5.

**(a) Emission intensity Image from a portion of grown MTSQDs**

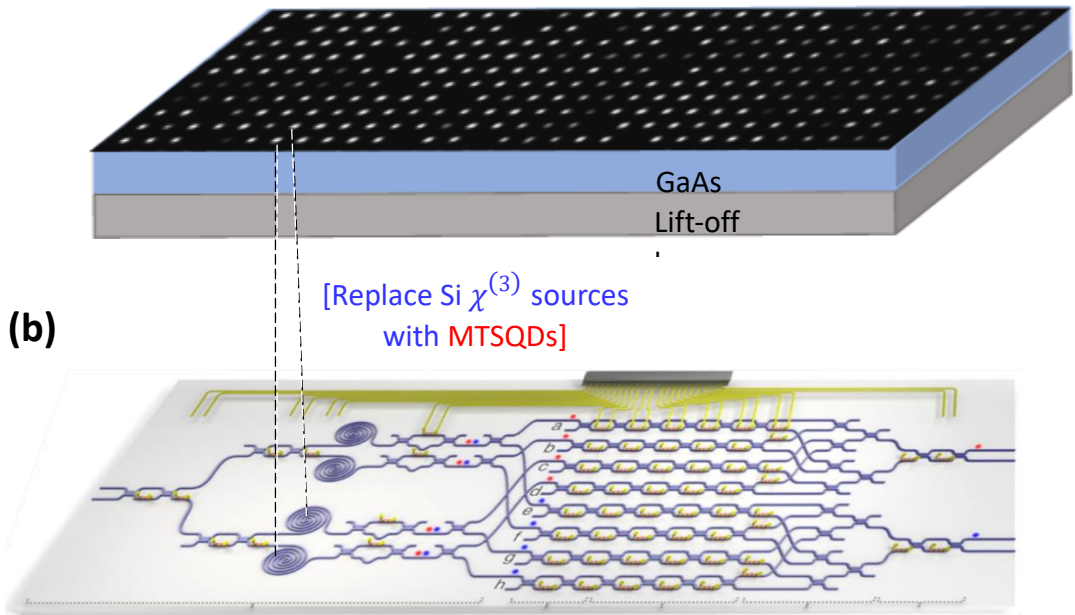

**Figure 5.** (a) A photon intensity image from a ~ 30 × 10 region of a 100 × 100 MTSQD array; Suitably co-designed arrays of MTSQDs can be hybrid integrated with (b) Si photonics platform based LOQC circuitry [from Ref. 16] to replace the spontaneous four-wave mixing based probabilistic photon sources by the on-demand high rate MTSQD sources.

Likewise, for quantum networks aimed at quantum communication leveraging the existing global fiber network in the ~1550nm C-band [17], quantum emitters, including quantum dots, emitting in the C-band are highly desirable otherwise an added feature of frequency conversion needs to be added to the chip. Other than the all-optical networks [18], long distance quantum networks demand quantum repeaters and consequently quantum memory at each node of the network [19].

Suitably designed epitaxial quantum dots thus are, in principle, capable of providing the required emission characteristics for the on-demand quantum emitters needed at desired wavelengths stretching from the short (790nm to 850nm, suited for space applications) to fiber communication (1300nm and 1550nm neighborhoods). Despite these well-known (and certain other relevant experimentally established) facts, the conversion of quantum optics to viable quantum photonic technological platforms has not progressed as rapidly. The basic missing link has been the lack of a methodology for fabricating designed *spatially regular arrays* of *scalable* quantum dots emitting in the desired range of tailored wavelengths *while maintaining, across the array, figures of merit satisfying the spectral emission requirements depicted in Figure 1.* This is the *necessary* requirement [1]. Here we have reported early results that indicate that the mesa-top single quantum dots are highly promising for providing this critically needed and long-awaited bridge from quantum optics to on-chip scalable quantum photonic information processing systems, as commented on next.

## IV. Outlook: MTSQD Arrays and On-Chip Integrated Scalable Quantum Photonic Platforms

The US National Academy of Science report [2] titled "Quantum Computing: Progress and Prospects (2019)" concluded (p.218):

"For quantum dots, a major limitation currently arises from the difficulties in developing well-controlled and reproducible fabrication methods: because the optical properties of a quantum dot depend on its size and shape, uniform and predictable quantum dot sizes are critical."

The optical properties of the shape and size controlled quantum emitters - in spatially designed arrays for MTSQDs - presented here are evidence that this limitation has largely been removed by the development of MTSQDs with reproducible adequate (Figure 1) spectral emission characteristics.

The basic platform of buried quantum emitters arranged in designed spatially-regular large arrays provided by the SESRE approach (Figure 2) enables the necessary deterministic embedding of these on-demand single photon emitters in *co-designed* on-chip fabricated light manipulating units (cavity, waveguide) for efficiently generating, from two or more quantum emitters chosen from a large set, photons directed into a waveguide mode for subsequent controlled interference and entanglement through an interconnected network representing the desired quantum optical circuits. The current [1] emission figures-of-merit [single photon purity > 99.5% and HOM visibility (extrapolated to 4K) ~90%] with ~3nm spectral standard deviation across large arrays of these alloy GaAs/InGaAs/GaAs mesa-top single quantum dots make a compelling *prima-facie* case for (1) continued studies aimed at improving the background quality of the material to minimize (potentially eliminate) spectral diffusion induced emission inhomogeneity; (2) optimizing grown structures (ideally avoiding alloys) and growth conditions for minimal inhomogeneity owing to the interfacial composition fluctuations controlling the statistical nature of the three-dimensional quantum confinement potential surface [10]; (3) incorporating the ability to locally fine-tune the emission to enable controlled interference and entanglement between photons generated in desired distinct quantum dots.

### Critical Role of Spatial Regularity

The unprecedented success of VLSI technology rests upon the ability to design and implement the physical locations of the basic building blocks comprising the transistor, resistor, capacitor, and inductor to reproducibly achieve interconnected network. This basic requirement of human-made architectures of on-chip information processing systems following the laws of classical physics remains even for the building blocks of on-chip information processing systems guided by the laws of quantum mechanics. Thus, until the time when human knowledge allows implementing architectures that mimic / emulate adequately the architecture of the naturally evolved information processing system—the brain—one must be able to place the building elements of the quantum information processing (QIP) system with adequate degree of accuracy and precision. Broadly viewed, a QIP system may be thought of as comprising five functional units interconnected in a seamless network to provide the desired overall primary function from the chip: (1) photon source unit: an on-demand single or entangled photon pair generator with appropriate characteristics. To obtain the desired coherence and collection efficiency of the single photons, the SPSs must be deterministically integrated with appropriate cavity + waveguide structures requiring ~nm scale precision in positioning. (2) an emitted photon manipulation subsystem (the network) containing interconnected nonlinear optical component for frequency conversion, if needed; (3) on-chip memory, if needed; (4) interconnected array of MZIs for photon state manipulation, and (5) appropriate detectors arranged as-designed to measure (compute / process) and transmit the information to the external classical system interfacing with humans. The transition from remarkable quantum optics studies and proof-of-principle demonstrations to workable quantum photonic technology is awaiting the improvement of a number of requisite characteristics of the building blocks but even with the current characteristics the ability to interconnect the building blocks into a functional system is thwarted to a very large part by the lack of a methodology for fabricating quantum emitters in spatially regular arrays with sufficient accuracy. Indeed, to date, the maximum number of on-chip quantum dot single photon sources monolithically integrated in a working network remains two [20]. With the non-scalable pick-and-place approach, about 100 SPSs have been integrated [21]. While ~100 is a remarkable number in the context of the past ~20 years of development, a road to scaling to larger systems is still not available. A revolution, equivalent to the VLSI in electronics, for on-chip quantum photonic systems requires a platform of SPSs in designed array with 1nm to 3nm precision-controlled placement over a large area. This work reports single photon emitters with adequate figures-of-merit across ~10,000 in an array realized using the SESRE approach that is readily scalable to ~1000 × 1000 (and beyond) arrays of MTSQDs providing millions of on-demand SPSs with the required ~nm scale precision in position, emission wavelengths within the local tunability range, *and* each emitter satisfying, with a specified high degree of statistical reliability, the required criteria of brightness, purity, and visibility (Figure 1) for implementing on-chip scalable quantum information processing systems. Arguably, this report constitutes a first step towards a scalable quantum photonic technology platform.

It is important to emphasize that the SESRE approach does not rely upon lattice mismatch induced strain to drive the formation of the mesa-top quantum dot—only the designed surface-curvature directed surface stress gradients and their contribution to the atomic-scale local surface energies / step-step interaction energies. Indeed, the first mesa-top quantum dots were realized in the lattice matched GaAs/AlGaAs system [22, 23]. Moreover, the recent remarkable demonstration [24] of near 0.1ms coherence time of electron spins achievable in GaAs/AlGaAs quantum dots (despite the nuclear spin bath) bodes well for the otherwise perceived weakness of the quantum dot platform, i.e., short matter (spin) qubit coherence time compared to the millisecond rare earth deep levels. Also, the SESRE approach is applicable to a variety of material combinations covering the wavelength regime from ultraviolet to mid and long IR regimes [1].

In closing, we emphasize that epitaxial quantum dots, through the efforts of many over the past two decades [4,5], have been demonstrated to inherently possess the required characteristics of single photons and polarization-entangled two-photon (via the biexciton-exciton decay cascade) pairs needed for use in quantum information processing (QIP) systems. There is no fundamental impediment to the MTSQDs exhibiting the best of the qualities demonstrated for other types of epitaxial quantum dots [5]. The SESRE approach places epitaxial quantum dots, of a vast combination of material systems, in designed spatially-ordered arrays with sufficient control on their location, size, and shape to allow the best of the established spectral characteristics to be produced in the large numbers necessary for a QIP system capable of addressing issues beyond the classical platforms. This is the uniqueness of the epitaxial quantum dots amongst all solid-state emitters even as others have their own strengths, primarily in the form of the entangled matter qubit coherence time being significantly longer. Ultimately, as is generally recognized, hybrid integrated on-chip systems involving trade-offs amongst the "best" of each to adequate and compatible with each other in implementable "workable" QIP systems that serve specific niches is what is the common goal.

## Appendix:  The Mesa-Top Single Quantum Dot Array Status

The large area spatially-ordered arrays of MTSQDs of Figure 2 are realized using the spatially-selective epitaxial growth technique of substrate-encoded size-reducing epitaxy (SESRE) introduced [22, 23] and developed [25, 26] at the Nanostructure Materials and Devices laboratory (NMDL). Briefly, SESRE creates quantum dots near the top of in-situ growth-reduced nanomesas (Figure 6(b), red region) fabricated in designed spatial arrays using nano lithography. The mesa edge orientation (Figure 6(a)) is chosen such as to induce adatom migration from the side facet to the top with negligible atom incorporation on the side facets to realize growth-controlled reduction of mesa top lateral size and hence the selective formation of the quantum dot on the in-situ size-reduced mesa top of prespecified lateral size (Figure 6(b), red region). Hence the name mesa top single quantum dots [26].

However, such *as-formed* single or vertically stacked multiple quantum dots are in a pillar-like pyramidal morphology unsuited for on-chip quantum optical circuits that require in-plane (horizontal) emission and propagation of the emitted photons and their interference and entanglement in horizontal networks. To meet this requirement, a significant advance introduced more recently [10] is the choice of the starting nanomesa shape (Figure 6(a)). The truncated pedestal shape (Figure 6(a)), we demonstrated, reverses, with continued growth past the pyramidal shape pinches-off, the relative incorporation rate between the valley and mesa top regions, thereby enabling planarization of the growth front morphology [10], resulting in a buried array of single quantum dots (Figure 6(c)). These provide the needed quantum emitter arrays for subsequent fabrication of optical quantum circuits. Figure 6 panels (d) and (e) show high-angle annular dark field (HAADF) scanning transmission electron microscope (STEM) based images [1] revealing, via the placement of AlAs (dark lines) marker layers during growth of GaAs (grey regions), the evolution of the growth front profile: initially the as-patterned size reduction, then the formation of the typical MTSQD, i.e. the placement of InGaAs (white region, panel (e)) followed side facet by pinch-off; then with  continued growth of GaAs the planarization of the growth front morphology. The spatial distribution of Ga, Al, and In was simultaneously investigated using energy dispersive spectroscopy (EDS). Panel (f) shows the EDS based chemical composition image revealing the quantum confined region dominated by In (violet region) surrounded by the Ga (blue region) of GaAs and the Al (green lines) of the AlAs marker layers.The HAADF and EDS results on the MTSQDs confirm the formation location, size, shape and composition following the SESRE growth method. The burying of the InGaAs regions in GaAs matrix with a flat surface morphology (i.e. the planarization overgrowth) enables the fabrication, through appropriate patterning and etching, of the designed photon manipulating unit around each MTSQD comprising photonic components such as a cavity, connected to efficiently collecting waveguide, onto directional couplers, etc. as needed for the designed functional quantum photonic network such as in Figure 5(b).

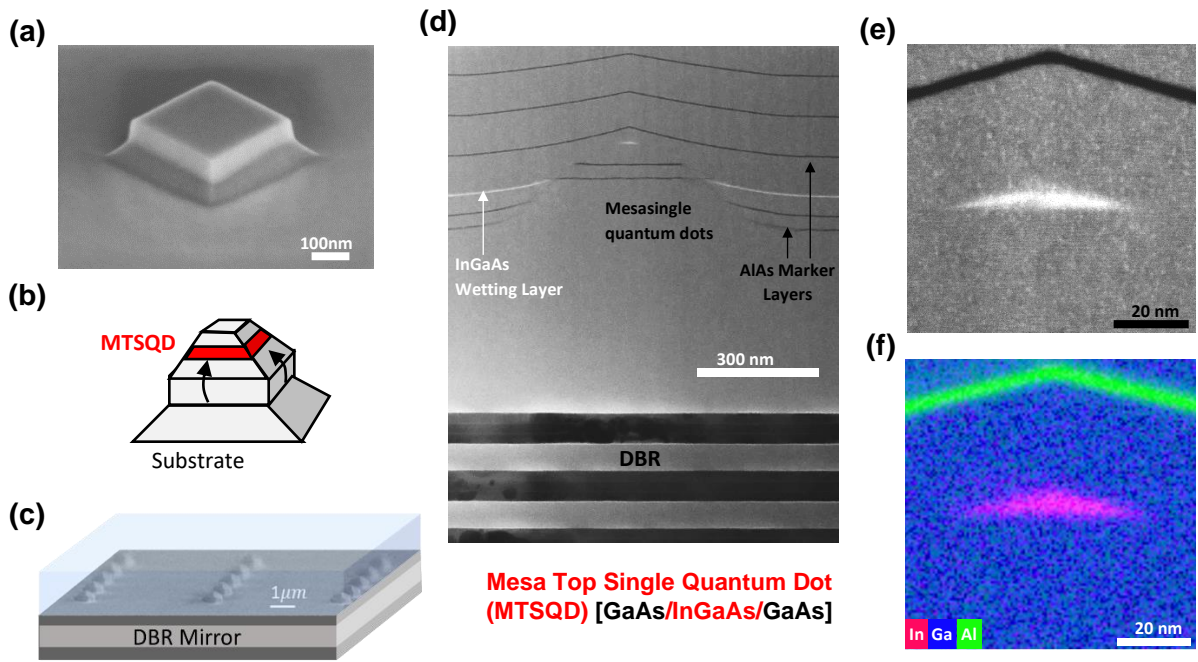

**Figure 6.** (a) SEM image of an as-fabricated pyramidal nanomesa suited for the growth of (b) schematic of single quantum dots (red region) formed via the SESRE approach on size-reduced top of nanomesas in arrays. (c) Schematic composite of an array of the *as-formed* pyramidal structures bearing MTSQD (the grey SEM image) buried in the GaAs planarized overlayer (shown translucent). (d) STEM (scanning transmission electron microscope) Z-contrast image indicating the evolution of the as-patterned GaAs (grey regions) mesa profile (dark lines) introduced through judicious placement of 10ML AlAs marker layers (dark lines). Note the light grey region just below the pinched-off marker layer—shown magnified in panel (e) is the region of the InGaAs quantum dot. Panel (f) shows the chemical composition revealed by EDS (energy dispersive spectroscopy) in STEM. The In (violet region) is seen to be confined to the expected region of the quantum dot. Blue and green denote GaAs and AlAs regions, respectively. Figure adapted from ref. 1.

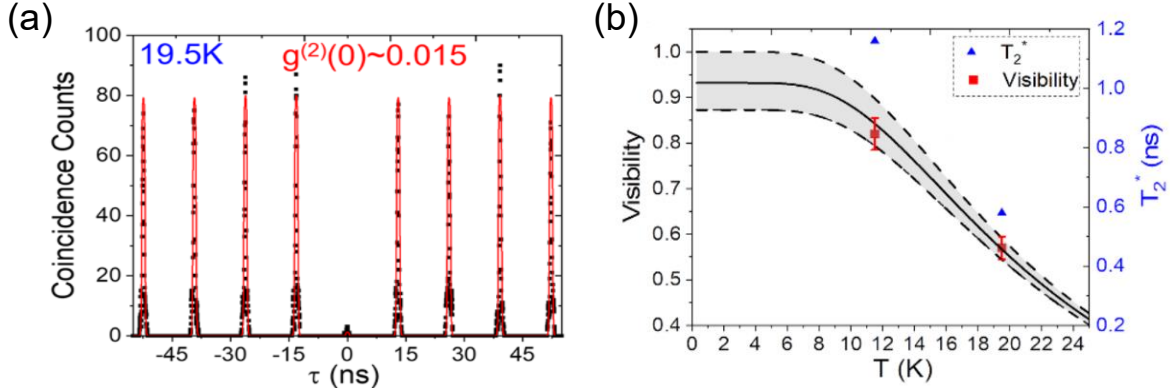

**Figure 7.** Taken from [ref. 1], panel (a) shows the typical HBT g²(0) at ~20K and panel (b) the HOM two-photon visibility at ~20K and at ~12K. The former indicates a single photon purity >99% and the latter, phonon broadening limited ~82%, extrapolates to better than 95% at the usual low temperature of ~4K at which most literatures values have been reported. Note that the reported values are with *no Purcell enhancement.* Taken from ref.1.

We have reported the typical basic optical characteristics of above shown GaAs/In$_{0.5}$Ga$_{0.5}$As/GaAs MTSQDs grown initially in 5×8 arrays [1, 25, 26] and scaled up to 10×10, 50×50, and 100×100. As a ready reference, an illustrative behavior is reproduced in Figure 7. Our cryogen-free cryostat is limited to ~19.5K and liquid helium being prohibitively expensive, currently we are unable to examine our samples at the typical~4K temperature at which much of the quantum dot literature is reported. Shown in Figure 7(a) is the typical HBT (Hanbury Brown-Twiss) coincidence count as a function of delay time between the detectors, g$^{(2)}$(τ) [1]. It indicates a single photon purity >99%. Figure 7(b) shows the behavior of the photon indistinguishability as measured by the HOM (Hong-Ou-Mandel) two-photon interferometry [1]. As we discussed in ref.1, the visibility is largely limited by the phonon scattering at the elevated temperatures of the measurements and extrapolated to liquid helium temperature based upon well-tested theoretical analysis, will lie above 90% even for the material quality of these samples.

The optical characteristics captured in Figure 7, combined with the time-resolved PL decay lifetime of ~0.35ns [1], clearly demonstrate that MTSQDs satisfy the requirements demanded of quantum emitters as shown in Figure 1. The nearly a third spontaneous decay time of the neutral excitons of MTSQDs of Figure 7 compared to the typical ~1ns for the self-assembled quantum dots SAQDs employed in the vast majority of the epitaxial SQD literature is a consequence of the significantly higher electric dipole moment of ~70Debye for the "size and shape" (i.e. the effective confinement potential) of these MTSQDs [1]. Indeed, with optimization, the oscillator strengths can be further enhanced [11]. To improve on the current numerical values of the figures-of-merit, needed foremost is improved material quality—largely growth in an MBE chamber of higher background purity and using the highest purity starting source materials with finely optimized growth protocol (growth parameters and procedure). It is worth noting that as with the SAQDs [5], the MTSQDs provide negatively or positively singly-charged excitons (i.e. trions) that offer single photon source without the fine structure splitting typical to the neutral exciton owing to loss of appropriate symmetry. Moreover, the biexciton-exciton decay cascade offers polarization entangled photons that serve as Bell-pair generation resource [5]. Note that these are not heralded photon pairs as the two are emitted at two different random times. The s*patially-ordered* MTSQDs can thus replace the randomly located and highly inhomogeneously emitting SAQDs in various conceived QD-based on-chip systems but which cannot be implemented with SAQDs.

It is important to emphasize that SESRE is equally effective for creating SQDs of lattice-matched (e.g. GaAs/AlGaAs) and lattice mismatched (e.g. GaAs/In$_x$Ga$_{1-x}$As with x up to 1) material combinations. Their single photon emitter nature has been established [10, 26]. The MTSQD arrays thus bridge the existing gap between quantum optics and realizing on-chip integrated and scalable quantum photonic platforms for technological applications. The nature of our growth chamber constrains the growth to the AlGaInAs material system, thereby limiting the emission wavelength regime to basically from ~750 nm (GaAs/AlAs) to 1300nm (GaAs/InAs) although with intervening strain-relieving layers incorporated [27], GaAs/In$_x$Ga$_{1-x}$As/InAs/ In$_y$Ga$_{1-y}$As/GaAs quantum dots can be created that emit near 1550nm. Of course, SESRE can be implemented for other material combinations within the III-V semiconductor family (Sb, P) or other classes of materials (II-VI, TMDCs, etc.) to cover extended wavelength regimes, including the C-band (~1550nm) made technologically important for fiber-based quantum communication given the existing classical communication infrastructure.

**Acknowledgements**: This work is supported by the Air Force Office of Scientific Research [PM Dr. Gernot Pomrenke] Grants FA9550-17-1-0353 and FA9550-22-1-0376, the Center for Nanoimaging (CNI) at the University of Southern California, and the Kenneth T. Norris Professorship.